\documentclass[10pt]{article}
\usepackage[square,authoryear]{natbib}
\usepackage{marsden_article}
\usepackage{epstopdf}
\usepackage{setspace}
\usepackage{bbm}

\usepackage{amscd}

\usepackage{caption2}

\DeclareGraphicsExtensions{.eps,.ps,.pdf}

\begin{document}

 \newtheorem{thm}{Theorem}[section]
 \newtheorem{cor}[thm]{Corollary}
 \newtheorem{lem}[thm]{Lemma}
 \newtheorem{prop}[thm]{Proposition}
 \newtheorem{defn}[thm]{Definition}
 \newtheorem{rem}[thm]{Remark}
 \numberwithin{equation}{section}

\title{\Large{\textbf{Effect of strain and oxygen vacancies on the structure
of\\ $\mathbf{180^{\circ}}$ ferroelectric domain walls in PbTiO$_3$}}}

\author{Arzhang Angoshtari\thanks{School of Civil and Environmental Engineering,
  Georgia Institute of Technology, Atlanta, GA 30332}
  \and  Arash Yavari\thanks{School of Civil and Environmental Engineering,
  Georgia Institute of Technology, Atlanta, GA 30332. E-mail: arash.yavari@ce.gatech.edu.}
   }

\maketitle

\begin{abstract}
In this paper we study the effect of normal and shear strains and
oxygen vacancies on the structure of $180^{\circ}$ ferroelectric
domain walls in PbTiO$_3$. It is known that oxygen vacancies move to
the domain walls and pin them. Hence, we assume a periodic
arrangement of oxygen vacancies on both Pb-centered and Ti-centered
domain walls in PbTiO$_3$. We use a semi-analytic anharmonic lattice
statics method for obtaining the relaxed configurations using a
shell potential. In agreement with recent ab initio calculations, we
observe that a Pb-centered domain wall with oxygen vacancies is not
stable even under strain. Our semi-analytic calculations for
PbTiO$_3$ show that oxygen vacancies affect the structure of
$180^{\circ}$ domain walls significantly but do not have a
considerable effect on the thickness of domain walls; they broaden
the domain walls by about fifty percent. We also study the effect of
normal and shear strains on both perfect and defective $180^{\circ}$
domain walls. We observe that normal and shear strains affect the structure but do not change the domain wall thickness.
\end{abstract}





\section{Introduction}

Ferroelectric perovskites have been the focus of intense research
in recent years because of their potential applications in high
strain actuators, high density storage devices, etc.
\citep{BhattacharyaRavichandran2003}. It is known that macroscopic
properties of ferroelectrics are strongly dependent on domain
walls, which are extended two-dimensional defects. Any fundamental
understanding of ferroelectricity in perovskites requires a detailed
understanding of domain walls in the nanoscale as these defects
are atomically sharp (see \cite{DawberRabeScott2005} and
references therein). Theoretical studies of domain walls have
revealed many of their interesting features. From both ab inito
calculations \citep{MeyerVanderbilt2001,Padilla1996,Poy1999,Poy2000} and anaharmonic lattice statics calculations
\citep{YaOrBh2006b} it is now known that ferroelectric domain
walls are atomically sharp. In all these studies, the structure
calculations have been done for perfect domain walls and free of strain. However,
domain walls interact with other types of defects and mainly with
oxygen vacancies and this may affect the structure and properties
of domain walls. Strain may also have a significant effect on domain wall structure.

It is known that the presence of point defects can have important
effects on the properties of perovskites. For example,
\citet{Bujakiewicz-KoronskaNatanzon2009} show that point defects
alter the elastic constants of Na$_{1/2}$Bi$_{1/2}$TiO$_3$
significantly. In this paper, we study the effect of oxygen
vacancies and normal and shear strains on the structure of
$180^{\circ}$ ferroelectric domain walls. It is known that oxygen
vacancies move to the domain walls and pin them and thus we assume a
periodic arrangement of oxygen vacancies on both Pb-centered and
Ti-centered domain walls. We use a semi-analytic anharmonic lattice
statics method for obtaining the relaxed configurations using a
shell potential. In agreement with recent ab initio calculations, we
observe that a Pb-centered domain wall with oxygen vacancies is not
stable even under strain. Our semi-analytic solutions for PbTiO$_3$
show that oxygen vacancies increase the thickness of the domain wall
by about fifty percent. This is different from the results of a
recent experimental measurements of $90^{\circ}$ domain walls in
PbTiO$_3$ using AFM by \citet{Shilo2004}. They observed that there
is a large variation in domain wall thickness with respect to
position ($0.5-4.0$ nm). This is not very surprising as vacancies
may interact differently with different types of domain walls. In
our calculations, we observe that oxygen vacancies have a
significant effect on the detailed structure of $180^{\circ}$ domain
walls. Our result is in agreement with a recent continuum study of
interaction of oxygen vacancies with domain walls
\citep{XiaoShenoyBhattacharya2005}. Xiao, et al.'s continuum model
predicts that $180^{\circ}$ and $90^{\circ}$ domain walls have quite
different interactions with oxygen vacancies.

There are several works in the literature on the effect of
strain on ferroelectric domain walls showing that strain can
have important effects on domain walls. For example, both experiments
\citep{ShangTan2001} and ab initio calculations
\citep{Shimada2008II} show that shear stress applied to
$90^{\circ}$ domain walls in PbTiO$_3$ develops polarization
reorientation through a domain wall movement perpendicular to itself
known as stress-induced domain switching. To our best knowledge, there are no atomistic calculations in the literature on the effect of strain on the structure of $180^{\circ}$ domain walls in PbTiO$_3$.

 This paper is structured as follows. In \S 2 we briefly
review the previous studies of ferroelectric domain walls. We then
present the main ideas of anharmonic lattice statics analysis of
perfect and defective ferroelectric domain walls under strain. In \S 3 we report
some numerical results for $180^{\circ}$ domain walls in PbTiO$_3$
using a shell potential. Conclusions are given in \S 4.

\section{\textbf{Ferroelectric Domain Walls}}

Ferroelectric domain walls have been studied extensively both
theoretically and experimentally. For recent experimental
investigations see \citep{Shilo2004,
FranckRavichandranBhattacharya2006} and references therein.
\cite{Shilo2004} studied the structure of $90^{\circ}$ domain
walls in PbTiO$_3$ using atomic force microscopy (AFM). They
measured the surface topography of a given sample and compared it
with a displacement field that is obtained from the
Devonshire-Ginzburg-Landau phenomenological model. They calculated
the thickness of two domain walls at two different positions and
observed thicknesses of $0.5$ nm and $4.0$ nm for the two domain
walls. They then conjectured that presence of point defects is
responsible for this variation. \cite{Lee2005} used a two-dimensional square lattice model that
has a continuum order parameter interacting with a lattice of
Ising spins, where the Ising spins model the presence or absence
of point defects. They show that depending on the parameters used
in their model, one can reproduce the main features of the
experimentally observed variation in domain wall thickness in
\cite{Shilo2004}'s experiments. However, one should note that
Shilo, et al.'s results may not apply to all types of domain walls
as they studied only the $90^{\circ}$ domain walls.

Ab initio calculations of \cite{HeVanderbilt2003} show that oxygen
vacancies have a tendency to move to domain walls and pin them. They
also observed that defective domain walls are Ti-centered. In their
calculations, they had to assume a periodic array of domain walls
with a high density of charge-neutral oxygen vacancies. There have
also been other ab initio calculations of point defects in PbTiO$_3$
in the bulk \citep{Padilla1996, Park1998, Cockayne2004}.
\cite{XiaoShenoyBhattacharya2005} studied the effect of oxygen
vacancies on the structure of domain walls in tetragonal BaTiO$_3$
using a continuum theory that takes into account the fact that
ferroelectrics are wide-band-gap semiconductors. In their numerical
calculations they observed that $180^{\circ}$ and $90^{\circ}$
domain walls behave differently in response to oxygen vacancies. In
particular, they saw charge accumulation near $90^{\circ}$ domain
walls with a potential drop across the wall while these were absent
in the case of $180^{\circ}$ domain walls.

\paragraph{Anharmonic Lattice Statics of Domain Walls}

Method of lattice statics was introduced by \cite{Matsubara1952}
and \cite{Kanazaki1957} and was extensively used by Born and his
co-workers \citep{BoHu1988}. For more details and history see
\cite{OrtizPhillips2000, YaOrBh2006a, YaOrBh2006b}. In this paper, we study the possibility of domain wall broadening
by oxygen vacancies in the case of $180^{\circ}$ domain walls in
PbTiO$_3$. We also study the effect of shear and normal strains on both defect-free domain walls and domain walls with oxygen vacancies.

We present a semi-analytical solution of structure of
$180^{\circ}$ domain walls using an anharmonic lattice statics
method \citep{YaOrBh2006a, YaOrBh2006b}. From ab initio
calculations of oxygen vacancies in PbTiO$_3$
\citep{HeVanderbilt2003} and also molecular dynamics simulations
of CaTiO$_3$ \citep{Calleja1990}, we know that oxygen vacancies
have a tendency to move to the domain walls and pin them.
Therefore, we study the structure of domain walls with oxygen
vacancies sitting on the wall. To be able to solve the discrete
governing equations analytically we need to assume some
periodicity for the collection of vacancies on the wall. We
consider a $180^{\circ}$ domain wall in an infinite crystal and
assume that oxygen vacancies are periodically arranged on the
domain wall. Thus, the only restrictive assumption is the
periodicity and high density of vacancies on the domain wall. In
reality, oxygen vacancies can be distributed randomly and have a
smaller density. However, the results from our calculations can
still provide some important quantitative information on
the effect of oxygen vacancies on the structure of $180^{\circ}$
domain walls in the nanoscale.

\paragraph{Defect-Free $\mathbf{180^{o}}$ Domain Walls}

\begin{figure}[t]
\begin{center}
\includegraphics[scale=0.9,angle=0]{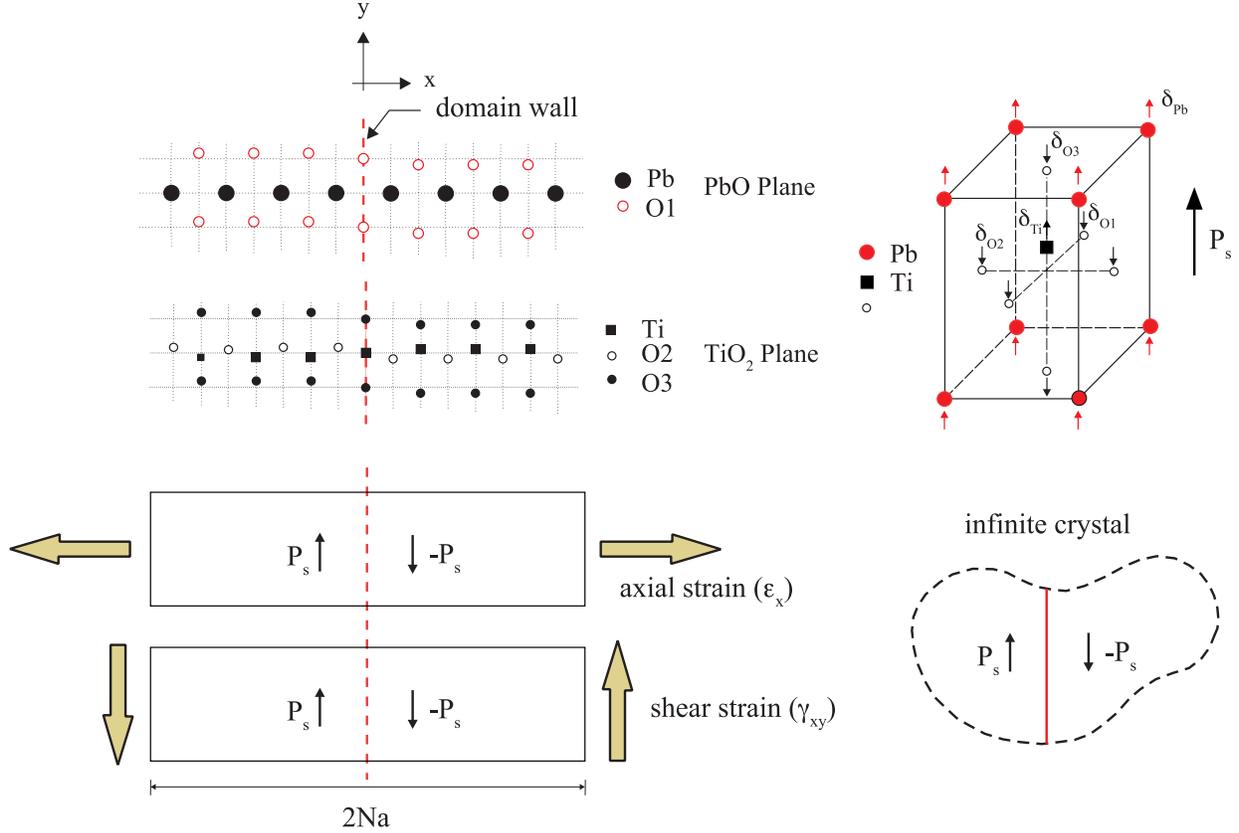}
\end{center}
\caption{\footnotesize Reference configuration for a Ti-centered
$180^{\circ}$ domain wall in PbO and TiO$_2$ planes. Note that cores and shells on the domain
wall have no relative shifts and cores and shells on the left and
right sides of the wall have opposite relative shifts. $a$ and $N$
are lattice parameter and size of the simulation box, respectively.
} \label{Reference}
\end{figure}

We use a shell potential for modeling PbTiO$_3$
\citep{Asthagiri2006} and all the calculations are performed for
$T=0$ K. Denoting the collection of cores and shells by
$\mathcal{L}$, $i\in\mathcal{L}$ represents a core (or shell) in the
collection. In a shell potential, total energy has the following
form \citep{DickOverhauser1964, Sepliarsky2000, Sepliarsky2004}
\begin{equation}
    \mathcal{E}\left(\left\{\mathbf{x}^i \right\}_{i\in \mathcal{L}}
    \right)=\mathcal{E}_{\textrm{short}}\left(\left\{\mathbf{x}^i \right\}_{i\in \mathcal{L}} \right)+\mathcal{E}_{\textrm{long}}\left(\left\{\mathbf{x}^i \right\}_{i\in
    \mathcal{L}} \right)+\mathcal{E}_{\textrm{core-shell}}\left(\left\{\mathbf{x}^i \right\}_{i\in \mathcal{L}}
    \right),
\end{equation}
where $\left\{\mathbf{x}^i \right\}_{i\in \mathcal{L}}$ is the
current position of cores and shells. Short range energy depends
explicitly on the position vectors of the massless shells. Long
range energy is the Coulombic energy of all cores and shells,
excluding core-shell interaction in the same atom. The
core-shell energy prevents the shell collapse in each atom and is
usually a polynomial function of the pairwise distance of core and
shell in a given atom. In the equilibrium configuration $\mathcal{B}=\left\{\mathbf{x}^i
\right\}_{i\in\mathcal{L}}\subset\mathbb{R}^3$, energy attains a
local minimum, i.e.
\begin{equation}\label{equilibrium}
    \frac{\partial \mathcal{E}}{\partial \mathbf{x}^i}=\mathbf{0}~~~~~~~\forall~i\in\mathcal{L}.
\end{equation}
However, in the case of a defective crystal, the problem is that
we do not know the relaxed configuration a priori. Thus, we start
with a reference configuration
$\mathcal{B}_0=\left\{\mathbf{x}_0^i \right\}_{i\in\mathcal{L}}$
that is not necessarily force free \citep{YaOrBh2006a}. In the
case of a $180^{\circ}$ domain wall, $\mathcal{B}_0$ is a nominal
defect, i.e., a configuration in which cores and shells on the
left and right sides of the wall have their positions in the bulk
configurations corresponding to $P_s$ and $-P_s$, respectively,
where $P_s$ is the spontaneous polarization (see Fig.
\ref{Reference}). This reference configuration is not force free. Let us denote the
discrete field of unbalanced forces by
\begin{equation}
    \mathbf{f}=\left\{-\frac{\partial\mathcal{E}}{\partial \mathbf{x}^i}\left(\mathcal{B}_0\right)\right\}_{i\in\mathcal{L}}.
\end{equation}
In anharmonic lattice statics one finds the discrete
deformation mapping $\varphi:\mathcal{B}_0\rightarrow\mathcal{B}$
that takes the chosen nominal defect to its relaxed configuration. Note that a
different choice of reference configuration $\mathcal{B}'_0$ leads
to a different discrete deformation mapping
$\varphi':\mathcal{B}'_0\rightarrow\mathcal{B}$. If the two
reference configurations are `close', i.e., they are in the same energy
well, we would converge to the same relaxed configuration. The discrete deformation map is found semi-analytically as follows.

Taylor expanding (\ref{equilibrium}) about the reference
configuration and ignoring terms higher than quadratic in
displacements, one obtains the following linearized governing
equations
\begin{equation}\label{equilibrium-linear}
    \frac{\partial\mathcal{E}}{\partial \mathbf{x}^i}\left(\mathcal{B}_0\right)+\sum_{j}\frac{\partial^2
    \mathcal{E}}{\partial\mathbf{x}^j\partial\mathbf{x}^i}\left(\mathcal{B}_0\right)(\mathbf{x}^j-\mathbf{x}_0^j)=\mathbf{0},
\end{equation}
or
\begin{equation}
    \sum_{j}\frac{\partial^2\mathcal{E}}{\partial\mathbf{x}^j\partial\mathbf{x}^i}\left(\mathcal{B}_0\right)\mathbf{u}^j=\mathbf{f}_i~~~~~~~~\forall~i\in\mathcal{L},
\end{equation}
where $\mathbf{u}^j=\mathbf{x}^j-\mathbf{x}_0^j$. For a
defect-free domain wall, atoms (cores, shells) of the same type
parallel to the wall will have the same displacement vectors. This
symmetry simplifies the linear equations
(\ref{equilibrium-linear}) considerably \citep{KavianpourYavari2009}.
Assume that the defective crystal $\mathcal{L}$ has a 1-D symmetry
reduction, i.e., it can be partitioned into two-dimensional
equivalence classes, i.e.
\begin{equation}
    \mathcal{L}=\bigsqcup_{\alpha\in\mathbb{Z}}\bigsqcup_{I=1}^{M}\mathcal{S}_{I\alpha}(i),
\end{equation}
where $\mathcal{S}_{I\alpha}(i)$ is the equivalence class of all the
atoms of type $I$ and index $\alpha$ with respect to atom $i$. Here
we assume that $\mathcal{L}$ is a multilattice of $M$ simple
lattices (for PbTiO$_3$, $M=10$). Particle $i$ is an arbitrary core (shell) and assuming that it is
in the n\emph{th} equivalence class of its type, $\mathcal{S}_{I\alpha(i)}$
is the set of those cores (shells) of type $I$ that are in the
equivalence class $n+\alpha$. For a free surface, for example, each
equivalence class is a set of cores (shells) lying on a plane
parallel to the free surface. Using this partitioning one can write
\begin{equation}
    \sum_{j}\frac{\partial^2 \mathcal{E}}{\partial \mathbf{x}^j\partial
    \mathbf{x}^i}\left(\mathcal{B}_0\right)(\mathbf{x}^j-\mathbf{x}^j_0)
    =\sum_{\alpha=-\infty}^{\infty}\sideset{}{'}\sum_{I=1}^M\sum_{j \in \mathcal{S}_{I\alpha(i)}}
    \frac{\partial^2 \mathcal{E}}{\partial \mathbf{x}^j \partial \mathbf{x}^{i}}(\mathcal{B}_0)
    \left(\mathbf{x}^{I\alpha}-\mathbf{x}^{I\alpha}_0\right),
\end{equation}
where the prime on the second sum means that the term $\alpha=0,
I=i$ is omitted. The linearized discrete governing equations are
written as
\begin{equation}
    \sum_{\alpha=-\infty}^{\infty}\sideset{}{'}\sum_{I=1}^{M} \mathbf{K}_{iI \alpha}
    \mathbf{u}^{I \alpha}+
    \left[-\!\!\sum_{\alpha=-\infty}^{\infty}\sideset{}{'}\sum_{I=1}^{M} \mathbf{K}_{iI \alpha}\right]\mathbf{u}^i=
    \mathbf{f}_i,
\end{equation}
where
\begin{equation}
    \mathbf{K}_{iI \alpha}=\sum_{j}\frac{\partial^2 \mathcal{E}}{\partial \mathbf{x}^j \partial \mathbf{x}^{i}}(\mathcal{B}_0),~~~
  \mathbf{f}_i=-\frac{\partial \mathcal{E}}{\partial \mathbf{x}^i}(\mathcal{B}_0),~~~
   \mathbf{u}^{I\alpha}=\mathbf{x}^{I\alpha}-\mathbf{x}^{I\alpha}_0=\mathbf{x}^{j}-\mathbf{x}^{j}_0
  ~~~~~ \forall ~j \in \mathcal{S}_{I\alpha(i)}.
\end{equation}
Unit cell displacement vectors are defined as
$\mathbf{X}_{n}=\left(\mathbf{u}^{1}_n,\hdots,\mathbf{u}^{M}_n\right)^{\mathsf{T}}$. Now the governing
equations in terms of unit cells displacements are
\begin{equation}\label{equilibrium-unit-cell}
    \sum_{\alpha=-m}^{m}\mathbf{A}_{\alpha}(n)\mathbf{X}_{n+\alpha}
    =\mathbf{F}_{n}~~~~~n\in\mathbb{Z},
\end{equation}
where
$\mathbf{A}_{\alpha}(n)\in\mathbb{R}^{3M\times3M},~\mathbf{X}_{n},
    \mathbf{F}_{n}\in\mathbb{R}^{3M}$ and
$m$ is the range of interaction of unit cells and $m=1$ would be
accurate enough for the shell potential \citep{YaOrBh2006b}.
Eq.(\ref{equilibrium-unit-cell}) is a linear vector-valued ordinary
difference equation of order $2m$ with variable coefficient matrices
and the unit cell force vectors and the unit cell stiffness matrices
are defined as
\begin{equation}
    \mathbf{F}_{n}=\left(
                       \begin{array}{c}
                         \mathbf{f}_{1n} \\
                         \vdots \\
                         \mathbf{f}_{Mn} \\
                       \end{array}
                     \right),~~~
    \mathbf{A}_{\alpha}(n)=\left(
                                            \begin{array}{cccc}
                                              \mathbf{K}_{11\alpha} & \mathbf{K}_{12\alpha} & \cdots & \mathbf{K}_{1M\alpha} \\
                                              \mathbf{K}_{21\alpha} & \mathbf{K}_{22\alpha} & \cdots & \mathbf{K}_{2M\alpha} \\
                                              \vdots & \vdots & \cdots & \vdots \\
                                              \mathbf{K}_{M1\alpha} & \mathbf{K}_{M2\alpha} & \cdots & \mathbf{K}_{MM\alpha} \\
                                            \end{array}
                                          \right)~~~~~n\in\mathbb{Z}.
\end{equation}
Note that $\mathbf{A}_{\alpha}(n)$ explicitly depends on $n$ and
this reflects the fact that close to the domain wall force
constants may change.

\paragraph{$\mathbf{180^{o}}$ Domain Walls With Oxygen Vacancies}

It is known for quite sometime that for a large enough number of
oxygen vacancies in perovskites one would see a self-organized
planar arrangement of oxygen vacancies (see \cite{ScottDawber2000,GopalanDierolfScrymgeour2007}
and references therein. See also \cite{Zhang2004} for atomistic
calculations and discussions on oxygen vacancies in BaTiO$_3$ and
their different possible arrangements). Therefore, similar to the
ab initio calculations of \cite{HeVanderbilt2003}, we assume that
oxygen vacancies interact with a domain wall and have a planar
structure coinciding with the domain wall.

\begin{figure}[t]
\begin{center}
\includegraphics[scale=0.7,angle=0]{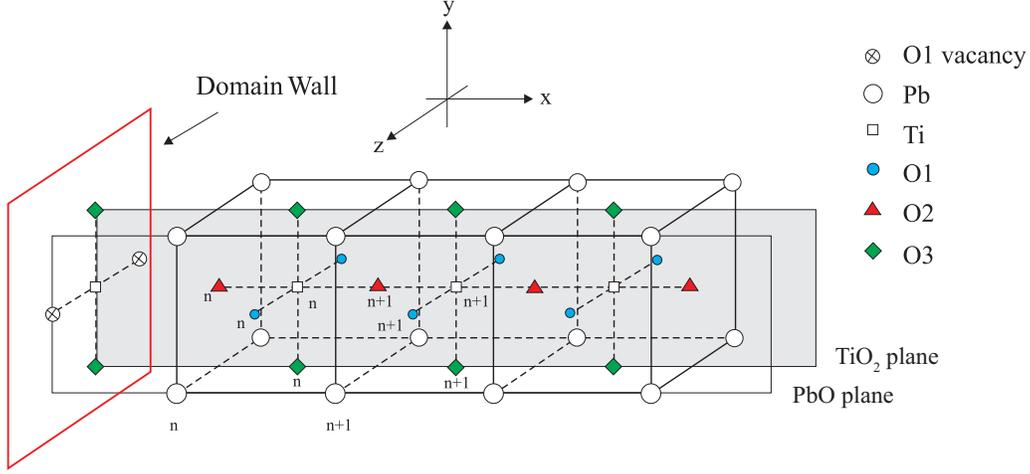}
\end{center}
\caption{\footnotesize Unit cell indexing in the reference
configuration of a Ti-centered $180^{\circ}$ domain wall. Note
that because of symmetry displacements in the z-direction are all
zero.} \label{Wall}
\end{figure}

Eq.(\ref{equilibrium-unit-cell}) is the governing equation
for a defective domain wall as long as the oxygen vacancies
are arranged periodically on the domain wall. However,
$\mathbf{A}_{\alpha}(n)$ and $\mathbf{F}_n$ would change, in general. We
can simplify the solution of the discrete boundary-value problem
even further by noting that the displacements of cores and shells
on the left side of the wall are related to those on the right
side of the wall, i.e., the wall is a reflection symmetry plane.
This reduces (\ref{equilibrium-unit-cell}) to an ordinary
difference equation on $\mathbb{N}_2$, i.e.,
\begin{equation}
    \mathbf{A}_{-1}~\mathbf{X}_{n-1}+\mathbf{A}_{0}~\mathbf{X}_{n}+\mathbf{A}_{1}~\mathbf{X}_{n+1}=\mathbf{F}_{n}~~~~~~~n\geq
    2,
\end{equation}
with the following boundary equations
\begin{eqnarray}
  && \mathbf{A}_{-1}(0)~\mathbf{X}_{-1}+\mathbf{A}_{1}(0)~\mathbf{X}_{1}=\mathbf{F}_{0}, \\
  && \mathbf{A}_{0}(1)~\mathbf{X}_{1}+\mathbf{A}_{1}(1)~\mathbf{X}_{2}=\mathbf{F}_{1}.
\end{eqnarray}
Note that $\mathbf{A}_{\alpha}(n)=\mathbf{A}_{\alpha},~n\geq 2$ and also $\mathbf{X}_{-n}=-\mathbf{X}_{n}~\forall n\geq 0$ and hence $\mathbf{X}_0=\mathbf{0}$. Note also that because of symmetry the z-component of all displacements are zero and hence $\mathbf{X}_n\in \mathbbm{R}^{20},~|n|\geq 1$. For a defective Ti-centered domain wall with O1 or O3 atoms removed, $\mathbf{X}_0\in \mathbbm{R}^4$ (see Fig. \ref{Wall}).

The governing discrete boundary-value problem is closed by
requiring boundedness of displacements at infinity, i.e.
\begin{equation}\label{finiteness}
    \lim_{n\rightarrow\infty}\parallel\mathbf{X}_{n}\parallel<\infty.
\end{equation}
One should note that the infinite crystal may undergo a rigid
translation after relaxation. Condition (\ref{finiteness}) does
not exclude this possibility. Note also that assuming that the
domain wall is a mirror plane for displacements removes the translation
invariance of the governing equations, i.e., those cores and
shells that lie one the domain wall in the reference configuration
remain on the domain wall after relaxation (i.e. $\mathbf{X}_0=\mathbf{0}$).

When there are oxygen vacancies on the domain wall, one needs to
look at boundary equations carefully. When modeling PbTiO$_3$ by a
shell potential, an oxygen vacancy means removal of the core and
shell of the oxygen atom and because we assume a charge neutral
oxygen vacancy, there will be a charge redistribution. This will
affect the shell charges of the neighboring shells.  It is known
that charge redistribution is highly localized. Thus, in our
calculations we distribute the charge $\Delta Q=Q_s+Q_c$, where
$Q_s$ and $Q_c$ are oxygen shell and core charges, only to the
(fourteen) first nearest neighbors of each oxygen vacancy. Most of
the existing shell potentials have fixed charges. However, in a more
realistic model shell charges should be variable to be able to
adjust themselves to the environment, e.g., in the presence of a
vacancy. There have been several efforts in the literature on
building empirical charge-variable models for different systems
based on Rappe and Goddard's charge equilibrium method
\citep{RappeGoddard1991} and its extensions. Here, we use a
fixed-charge shell potential for PbTiO$_3$ and are not aware of any
charge-variable shell model for this material. Therefore, we have to
model the charge transfer due to oxygen vacancies approximately.
However, this should not have severe effects on the results as our
numerical tests show. We studied the sensitivity of solutions to the
exact way of charge redistribution; we compared two cases: in the
first case we distributed the charge equally to the nearest
neighbors and in the second case we distributed the charge to
nearest neighbors depending on their distances from the vacancy and
assuming that charge distribution is exponentially decaying. We did
not see much difference and thus in this work we distribute the
charge equally between the nearest neighbors.

We consider three types of defective domain walls: (i) O2-defective,
(ii) O1-defective, and (iii) O3-defective. An O2-defective domain
wall is Pb-centered. See Fig. \ref{Reference} for this notation. We see
that the anharmonic lattice statics iterations do not converge in
this case, i.e., this is not a stable configuration\footnote{This is also the case when the defective domain wall is under strain.}. This is in
agreement with ab initio calculations \citep{HeVanderbilt2003}
that predict Ti-centered defective domain walls. O1 and O3-defective
domain walls are Ti-centered with no O1 and O3 cores and shells, respectively, on
the domain wall in the reference configuration (see
Fig. \ref{Wall}).

The linearized governing equations can be solved exactly using the
method proposed in \citep{YaOrBh2006a}. Thus, we are able to solve
the linearized governing equations exactly. Now to obtain the fully
nonlinear solutions we use a modified Newton-Raphson iteration.
Solving the linearized problem, we modify the reference
configuration by imposing the harmonic displacements and then
calculate the new unbalanced forces exactly using the interatomic
potential \citep{YaOrBh2006a}. Continuing in this manner, if there
is an equilibrium configuration close to the chosen reference
configuration, unbalanced forces converge to zero. In the present
work, convergence means that all forces have magnitudes less than
$0.05~eV/\AA$. Now, let us briefly explain the modified
Newton-Raphson method, called the quasi-Newton method, that we use
throughout this work.

Newton method is based on the following quadratic approximation near the
current configuration $\mathcal{B}^{k}$:
\begin{equation}
    \mathcal{E}\left(\mathcal{B}^{k}+\tilde{\boldsymbol{\delta}}^k\right)=\mathcal{E}\left(\mathcal{B}^{k}\right)
    +\nabla\mathcal{E}\left(\mathcal{B}^{k}\right)\cdot\tilde{\boldsymbol{\delta}}^k
    +\frac{1}{2}(\tilde{\boldsymbol{\delta}}^k)^{\textsf{T}}\cdot\mathbf{H}\left(\mathcal{B}^{k}\right)\cdot\tilde{\boldsymbol{\delta}}^k+o\left(|\tilde{\boldsymbol{\delta}}^k|^2\right),
\end{equation}
where
$\tilde{\boldsymbol{\delta}}^k=\mathcal{B}^{k+1}-\mathcal{B}^{k}$
and $\mathbf{H}$ is the Hessian matrix. By differentiating the above
formula with respect to $\tilde{\boldsymbol{\delta}}^k$, we obtain
the Newton method for determining the next configuration
$\mathcal{B}^{k+1}=\mathcal{B}^{k}+\tilde{\boldsymbol{\delta}}^k$:
\begin{equation}
    \tilde{\delta}^k=-\mathbf{H}^{-1}\left(\mathcal{B}^{k}\right)\cdot\nabla\mathcal{E}\left(\mathcal{B}^{k}\right).
\end{equation}
Note that in order to converge to a local minimum, the Hessian must
be positive definite.

If the calculation of the Hessian in each iteration becomes
numerically inefficient (like the present problem), one can use the
quasi-Newton method. The main idea behind this method is to start
from a positive-definite approximation to the inverse Hessian and to
modify this approximation in each iteration using the gradient
vector of that step. Close to the local minimum, the approximate
inverse Hessian approaches the true inverse Hessian and we would
have the quadratic convergence of the Newton method
\citep{PressTVF1989}. Here we use the
Broyden-Fletcher-Goldfarb-Shanno (BFGS) algorithm
\citep{PressTVF1989} for generating the approximate inverse Hessian:
\begin{equation}\label{BFGS}
    \mathbf{C}^{i+1}=\mathbf{C}^i + \frac{\tilde{\boldsymbol{\delta}}^k\otimes\tilde{\boldsymbol{\delta}}^k}{(\tilde{\boldsymbol{\delta}}^k)^{\textsf{T}}\cdot\mathbf{\Delta}}
    -\frac{\left(\mathbf{C}^{i}\cdot\mathbf{\Delta}\right)\otimes\left(\mathbf{C}^{i}\cdot\mathbf{\Delta}\right)}
          {\mathbf{\Delta}^{\textsf{T}}\cdot\mathbf{C}^{i}\cdot\mathbf{\Delta}}
    +\left(\mathbf{\Delta}^{\textsf{T}}\cdot\mathbf{C}^{i}\cdot\mathbf{\Delta}\right)\mathbf{u}\otimes\mathbf{u},
\end{equation}
where $\mathbf{C}^{i}=\left(\mathbf{H}^i\right)^{-1}$,
$\mathbf{\Delta}=\nabla\mathcal{E}^{i+1}-\nabla\mathcal{E}^{i}$,
and
\begin{eqnarray}
  \mathbf{u}=\frac{\tilde{\boldsymbol{\delta}}^k}{(\tilde{\boldsymbol{\delta}}^k)^{\textsf{T}}\cdot\mathbf{\Delta}}-
  \frac{\mathbf{C}^{i}\cdot\mathbf{\Delta}}{\mathbf{\Delta}^{\textsf{T}}\cdot\mathbf{C}^{i}\cdot\mathbf{\Delta}}.
\end{eqnarray}
Calculating $\mathbf{C}^{i+1}$, one then should use
$\mathbf{C}^{i+1}$ instead of $\mathbf{H}^{-1}$ to update the
current configuration for the next configuration
$\mathcal{B}^{k+1}=\mathcal{B}^{k}+\tilde{\boldsymbol{\delta}}^k$.
If $\mathbf{C}^{i+1}$ is a poor approximation, then one may need to
perform a linear search to refine $\mathcal{B}^{k+1}$ before
starting the next iteration \citep{PressTVF1989}.

\paragraph{$\mathbf{180^{o}}$ Domain Walls Under Strain}

We put both the perfect and defective domain walls under normal and
shear strains (see Fig. \ref{Reference}). We consider both
compressive and tensile strains and also shear strains both along
and opposite to the polarization directions. We apply strains to the
defective lattice by imposing displacements in the proper directions
far enough from the domain wall (displacements are in x- and
y-directions for normal and shear strains, respectively). Note that
the defective lattice is translated rigidly on both sides of the
domain wall outside the computational box. Applying strain to the
lattice should be done gradually. For both perfect and defective
domain walls, we start with the strain-free relaxed domain wall
configuration $\mathcal{B}$. Then, we apply proper boundary
displacement to increase (or decrease) strain by the value
$\Delta\epsilon$ to obtain the strained configuration
$\mathcal{B}_{\Delta\epsilon}$. Note that $\Delta\epsilon$ should be
small enough such that $\mathcal{B}$ and
$\mathcal{B}_{\Delta\epsilon}$ are close to each other. In the
present work, $\Delta\epsilon=0.001$ for normal strains and
$\Delta\epsilon=0.0005$ for shear strains. Next, we start with
$\mathcal{B}_{\Delta\epsilon}$ and calculate
$\mathcal{B}_{2\Delta\epsilon}$. Repeating this procedure, one can
apply large strains to the defective lattice and obtain its relaxed
configuration $\mathcal{B}_{\epsilon}$ for a given strain
$\epsilon$. Using the notation of the previous sections, we
summarize our algorithm for applying strain as follows.
 \vskip -0.6 in
\begin{picture}(150,150)
\framebox{\begin{minipage}[c]{3.5in}
\begin{itemize}
    \item [] Input data: $\mathcal{B}_\epsilon$, $\Delta \epsilon$ (strain increment)
    \item [$\vartriangleright$] Initialization
    \begin{itemize}
        \item [$\vartriangleright$] Apply B.C. to $\mathcal{B}_{\epsilon}$ and obtain $\mathcal{B}^1$
        \item [$\vartriangleright$] $\mathbf{H}^1=\mathbf{H}|_{\mathcal{B}=\mathcal{B}^1}$ and set $\mathbf{C}^1=(\mathbf{H}^1)^{-1}$
    \end{itemize}
    \item [$\vartriangleright$] Do until convergence is achieved
    \begin{itemize}
        \item [$\vartriangleright$] Calculate forces
        \item [$\vartriangleright$] Use quasi-Newton method to calculate
        $\mathbf{C}^{k+1}$
        \item [$\vartriangleright$] Use $\mathbf{C}^{k+1}$ to obtain
        $\mathcal{B}^{k+1}$
    \end{itemize}
    \item [$\vartriangleright$] End Do
    \item [$\vartriangleright$] End
\end{itemize}
\end{minipage}}
\end{picture}
\vskip 1.5 in

\section{Numerical Results}

\paragraph{Perfect Domain Walls}

In our numerical examples we show the anharmonic displacements with
respect to the reference configuration. The first step is to
calculate unbalanced forces. Our choice of reference configurations
makes the unbalanced forces localized in the direction perpendicular
to the domain wall. However, one should note that a domain wall is
an extended defect and unbalanced forces are not localized in the
tetragonal $c$-direction. We are able to handle this nonlocality
issue using the symmetry reduction idea. Since core and shell
displacements are close, we only report the core displacements in
this section. We assume $2N$ unit cells in the simulation box (see
Fig. \ref{Reference}). Our numerical experiments show that $N=10$
would be enough for calculating displacements as the structure does
not changed by using larger values for $N$.
\begin{figure}[t]
\begin{center}
\includegraphics[scale=0.7,angle=0]{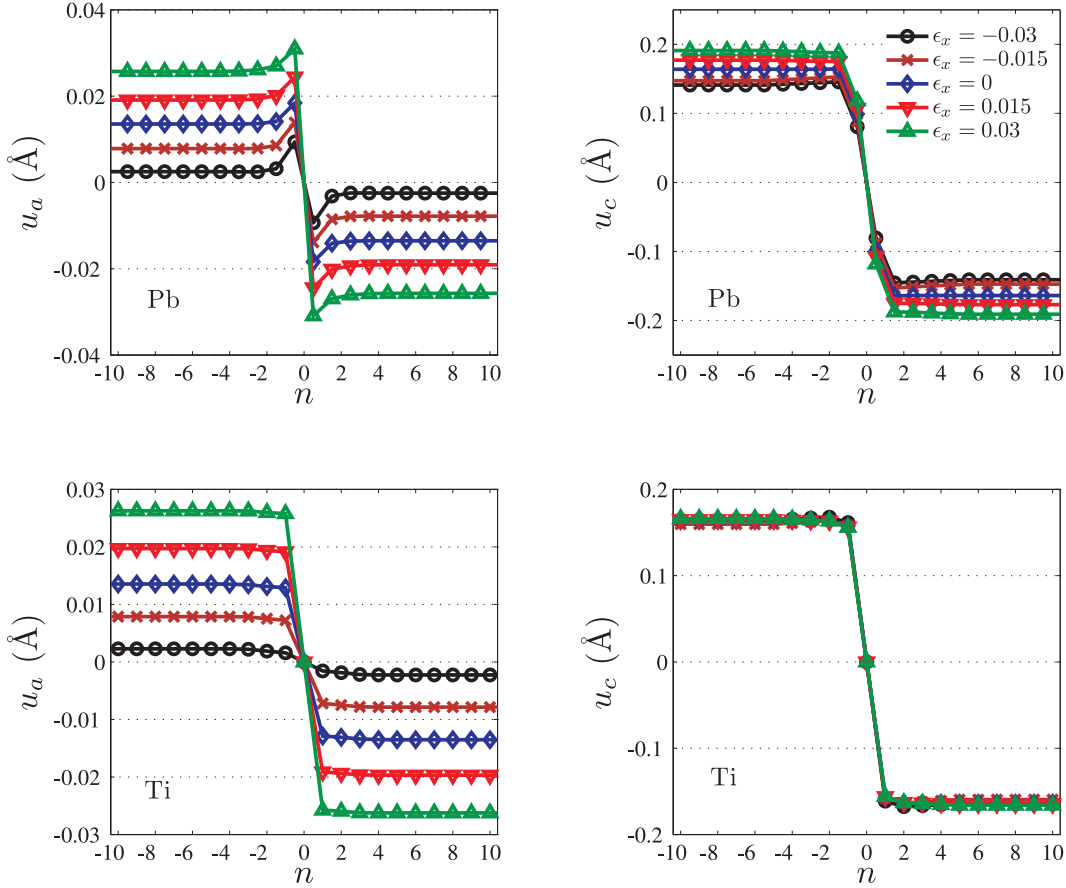}
\end{center}\vspace*{-0.0in}
\caption{\footnotesize $a$- and $c$-structure of a perfect
Ti-centered $180^{\circ}$ domain wall under axial strain
$\epsilon_x$. $u_a$ and $u_c$ are displacements along
$a$-direction and the tetragonal $c$-direction, respectively. }
\label{Ti_PT_ex}
\end{figure}
\begin{figure}[t]
\begin{center}
\includegraphics[scale=0.7,angle=0]{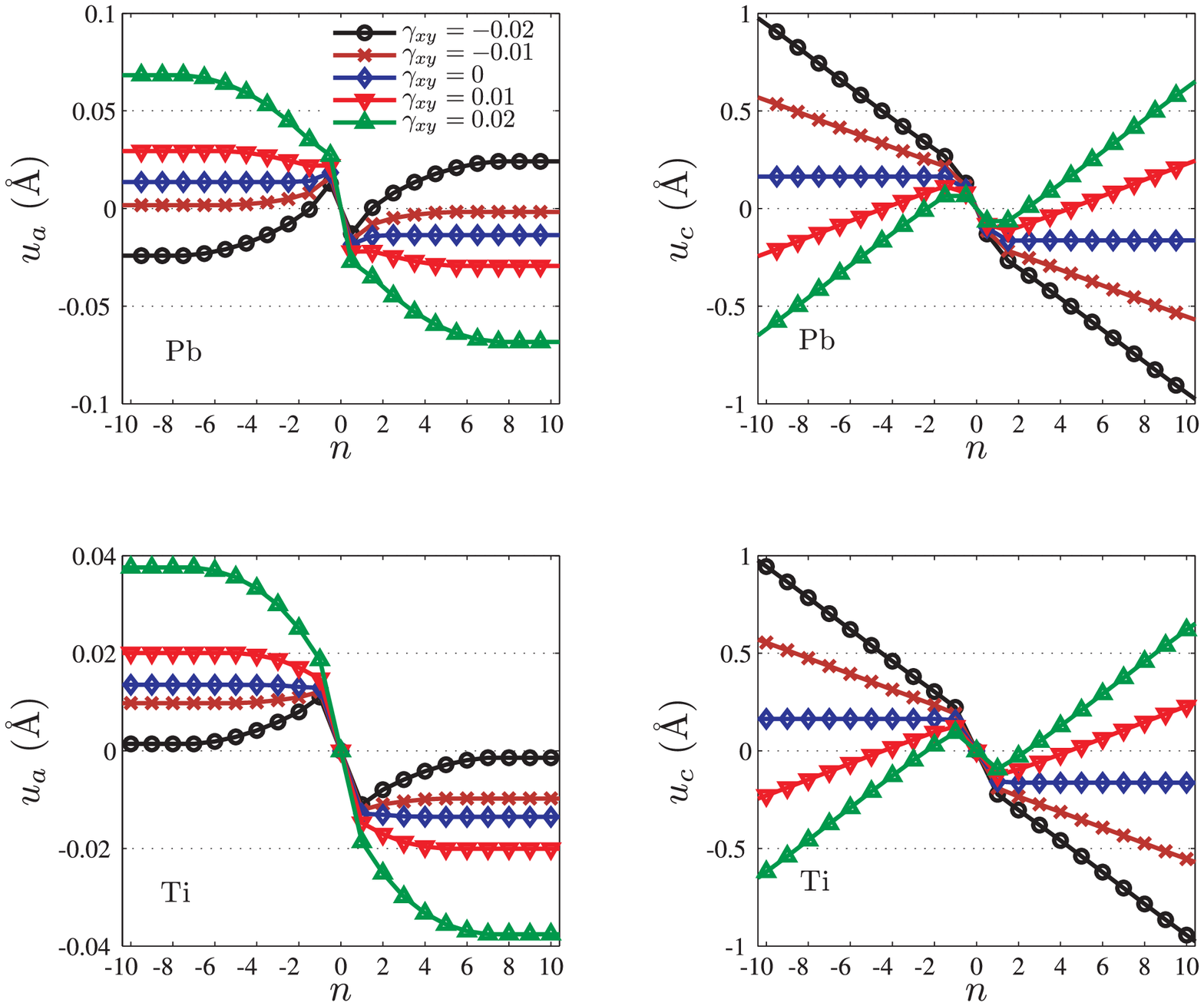}
\end{center}\vspace*{-0.0in}
\caption{\footnotesize $a$- and $c$-structure of a perfect
Ti-centered $180^{\circ}$ domain wall under shear strain
$\gamma_{xy}$. $u_a$ and $u_c$ are displacements along
$a$-direction and the tetragonal $c$-direction, respectively.}
\label{Ti_PT_gxy}
\end{figure}
\begin{figure}[h]
\begin{center}
\includegraphics[scale=0.7,angle=0]{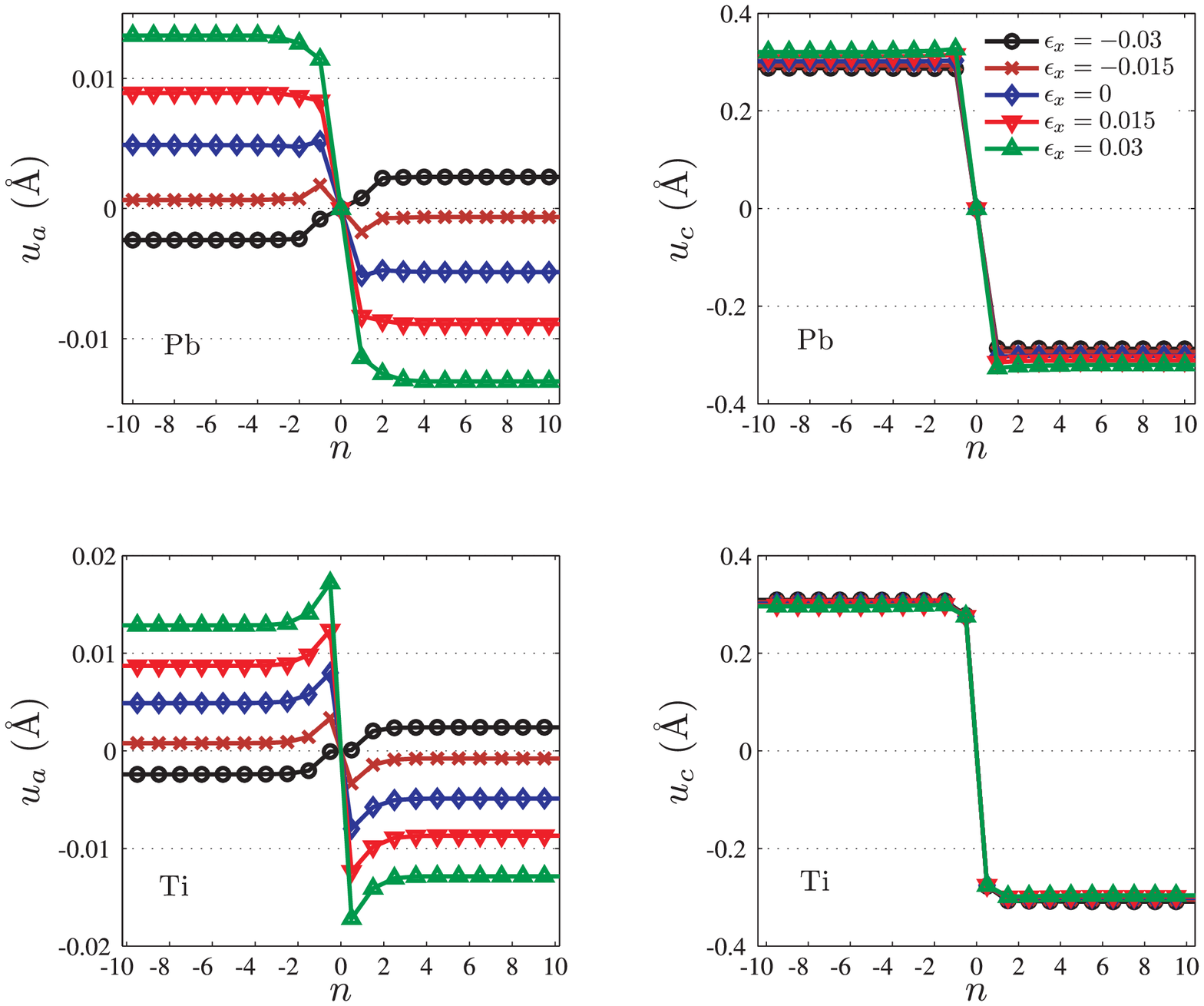}
\end{center}\vspace*{-0.0in}
\caption{\footnotesize $a$- and $c$-structure of a perfect
Pb-centered $180^{\circ}$ domain wall under axial strain
$\epsilon_x$. $u_a$ and $u_c$ are displacements along
$a$-direction and the tetragonal $c$-direction, respectively. }
\label{Pb_PT_ex}
\end{figure}

\begin{figure}[thb]
\begin{center}
\includegraphics[scale=0.7,angle=0]{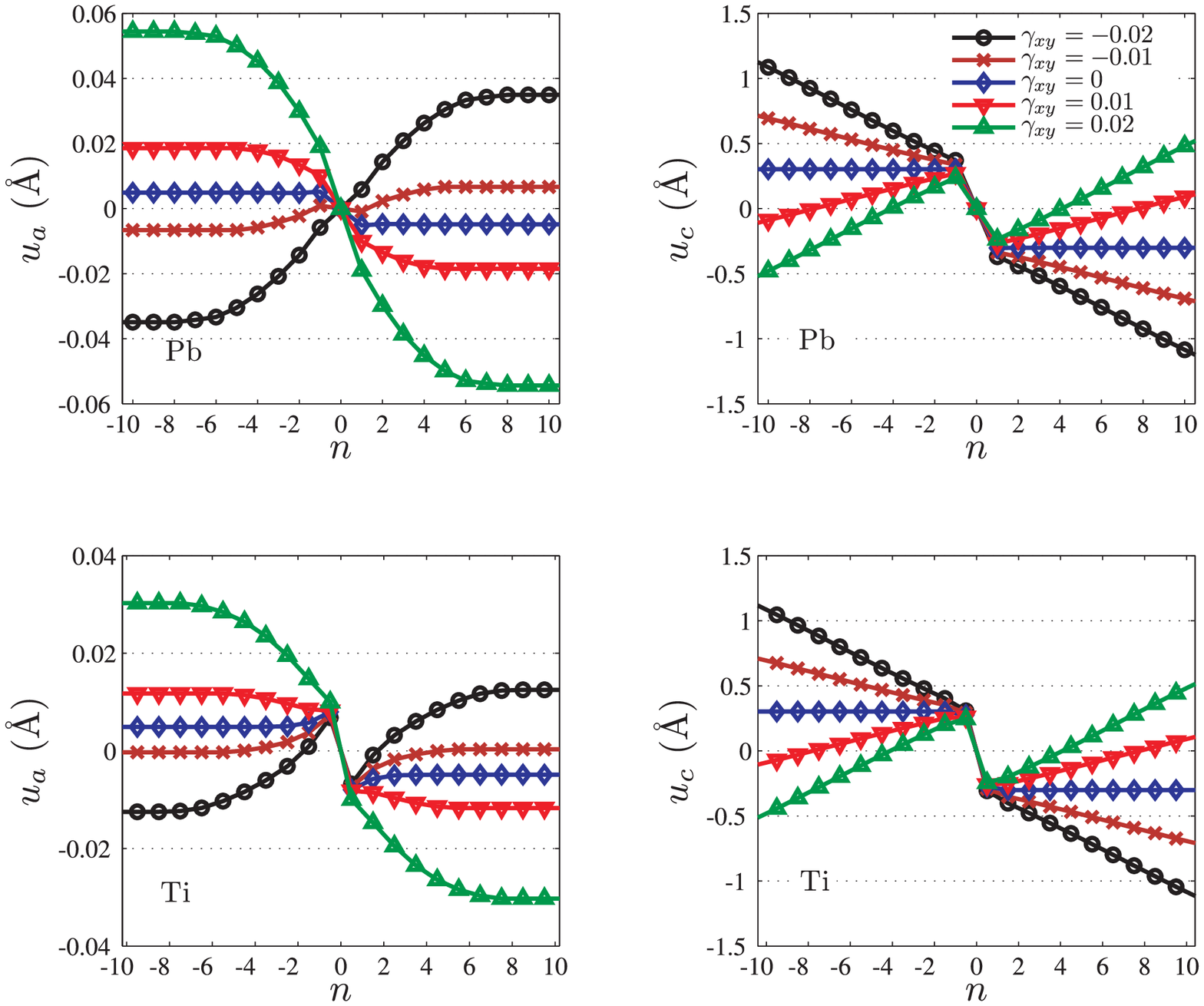}
\end{center}\vspace*{-0.0in}
\caption{\footnotesize $a$- and $c$-structure of a perfect
Pb-centered $180^{\circ}$ domain wall under shear strain
$\gamma_{xy}$. $u_a$ and $u_c$ are displacements along
$a$-direction and the tetragonal $c$-direction, respectively.}
\label{Pb_PT_gxy}
\end{figure}

Fig. \ref{Ti_PT_ex} shows the displacements of Pb and Ti cores in
the $a$- and $c$-directions ($u_a$ and $u_c$, respectively) for a
perfect Ti-centered domain wall under axial strain $\epsilon_x$. We
plot the displacements for different values of the axial strain
between $-0.03$ to $0.03$. As expected, $u_a$ has a larger
variation than $u_c$ under the axial strain. However, note that axial strain does not have a significant effect on domain wall
thickness; it is seen that all the distortions occur within two
lattice spacings on each side of the domain wall, i,e. domain wall thickness is about $1.0-1.5$ nm regardless of the value
of the axial strain. Fig. \ref{Ti_PT_gxy} shows displacements of the same domain wall
under different values of shear strain $\gamma_{xy}$. The
c-displacements are in the polarization direction for negative
values of $\gamma_{xy}$ and are in the opposite direction for
positive values of $\gamma_{xy}$. Hence, we do not see symmetric
displacements with respect to the unstrained structure. Also, we
observe that unlike axial strains, shear strains have a considerable
effect on both $u_a$ and $u_c$. Similar results for a perfect
Pb-centered domain wall are shown in Figs. \ref{Pb_PT_ex} and
\ref{Pb_PT_gxy}.

\paragraph{Domain Walls with Oxygen Vacancies}

\begin{figure}[thb]
\begin{center}
\includegraphics[scale=0.7,angle=0]{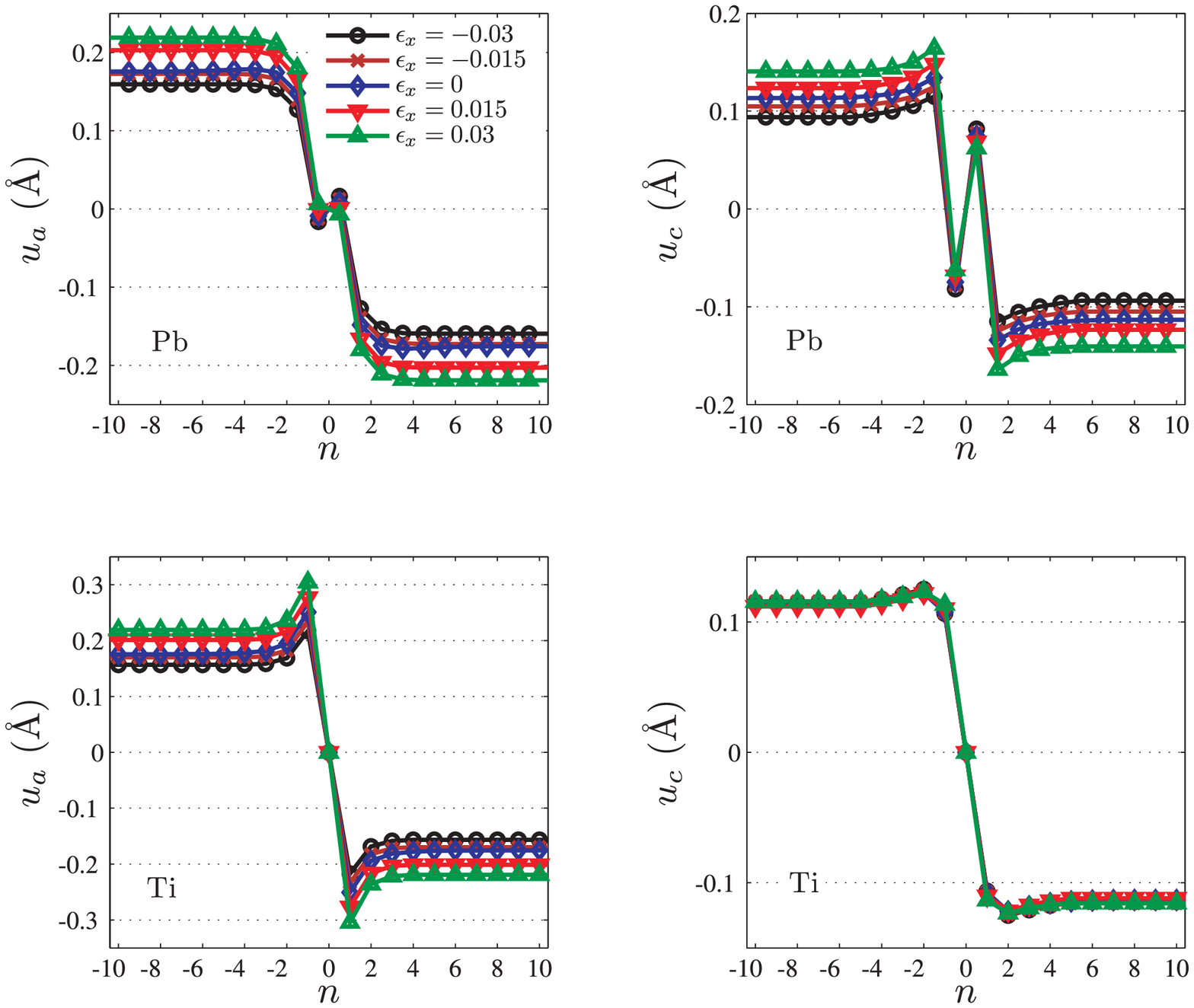}
\end{center}\vspace*{-0.0in}
\caption{\footnotesize $a$- and $c$-structure of an O1-defective
$180^{\circ}$ domain wall under axial strain $\epsilon_{x}$.
$u_a$ and $u_c$ are displacements along $a$-direction and the
tetragonal $c$-direction, respectively.} \label{O1Ti_PT_ex}
\end{figure}

We report the displacements of a Ti-centered $180^{\circ}$
domain wall with oxygen vacancies under axial and shear strains.
Since displacements of O1-defective and O3-defective walls are
similar, here we only report the results for O1-defective walls.
Fig. \ref{O1Ti_PT_ex} shows the displacements for an O1-defective
domain wall under normal strain. O1 vacancies lie on the domain wall
and because of symmetry have zero displacements, i.e. they will stay on the domain wall after deformation. For the strain-free
configuration, all distortions parallel to the domain wall
(c-displacements) occur within two lattice spacings on each side of
the wall, i.e., thickness of the domain wall in
c-direction is not affected by oxygen vacancies. However, it is seen
that structure is significantly different from that of a perfect
domain wall. It is also seen that unlike perfect domain walls,
$a$-displacements have the same order of magnitude as the
corresponding $c$-displacements. We observe that the $a$-displacements are
nonzero within three lattice spacings on each side of the wall.
Thus, the thickness of an O1-defective domain wall is about $1.5-2.0$
nm, i.e. oxygen vacancies increase the domain wall thickness by
about fifty percent. Here similar to perfect domain walls, we see that
normal strains do not have a significant effect on the thickness of
the domain wall.
\begin{figure}[thb]
\begin{center}
\includegraphics[scale=0.7,angle=0]{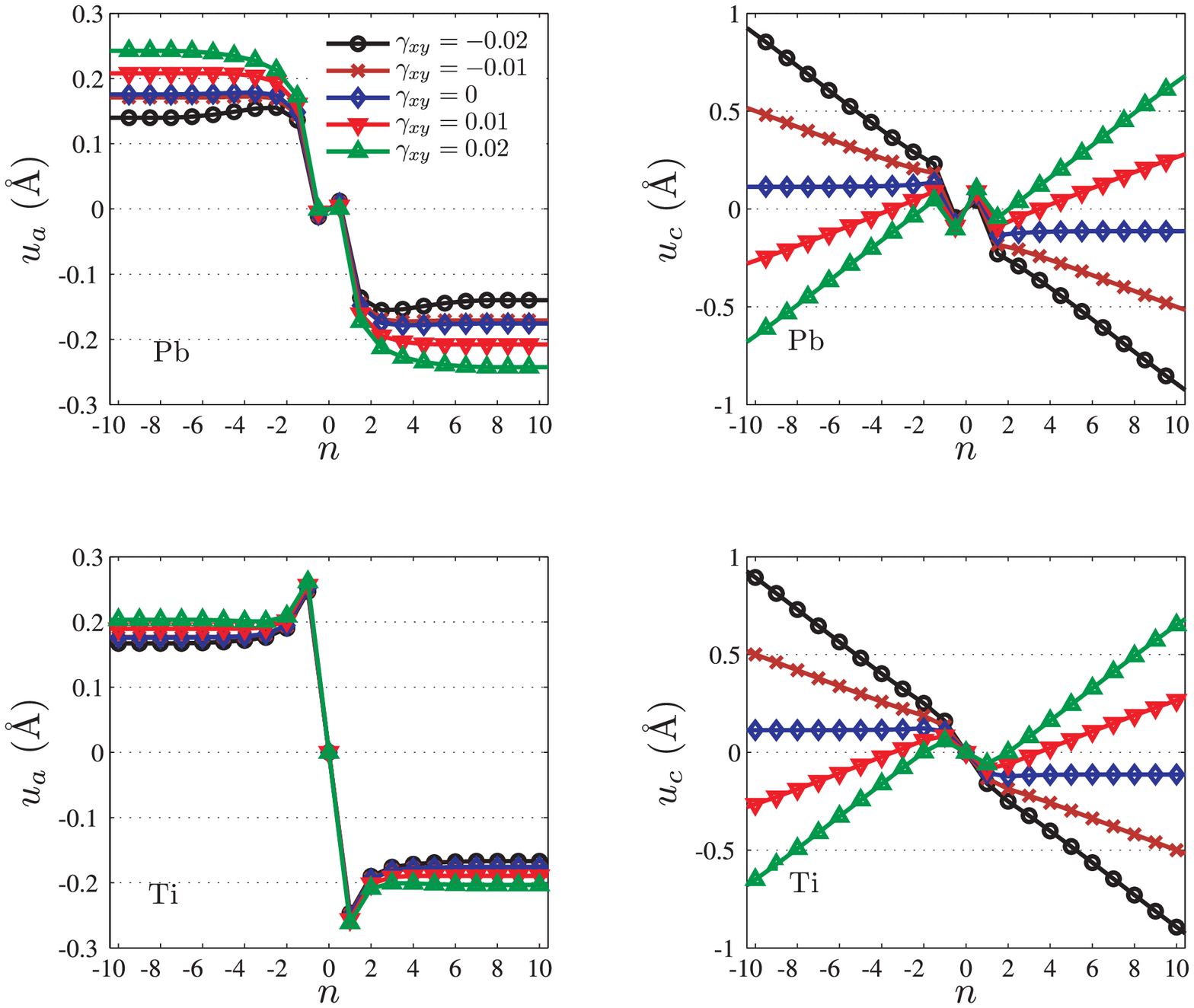}
\end{center}\vspace*{-0.0in}
\caption{\footnotesize $a$- and $c$-structure of an O1-defective
$180^{\circ}$ domain wall under shear strain $\gamma_{xy}$.
$u_a$ and $u_c$ are displacements along $a$-direction and the
tetragonal $c$-direction, respectively.} \label{O1Ti_PT_gxy}
\end{figure}
Fig. \ref{O1Ti_PT_gxy} shows the displacements of an O1-defective
domain wall under shear strain. Again, we see that shear strain has
a significant effect on the displacements of the domain wall. Also
note that domain wall thickness increases in $a$-direction, but this
increase is less than the increase observed for perfect wall (see
Fig. \ref{Ti_PT_gxy}).

The only restriction in our calculations is the high density of
oxygen vacancies (similar to the existing ab initio calulations). The
resulting stiffness coefficient matrices become highly
ill-conditioned by increasing the period of vacancies. Therefore, we
did the calculations only for the three cases where all O1, O2, or
O3 oxygen atoms are removed from the domain wall. In all the three
cases unbalanced forces in the tetragonal $c$-direction are nonzero
only in two unit cells on each side of the wall. We checked this for
several lower density arrangements of oxygen vacancies and observed
the same local property for unbalanced forces. In the case of
perfect domain walls, unbalanced forces perpendicular to the wall
are very small and nonzero only in two layers on each side of the
wall. In the case of defective domain walls, these unbalanced forces
are nonzero in three unit cells on each side of the wall. This was
also the case for several other lower density arrangements of oxygen
vacancies on the wall. Thus, we believe that the high density of
oxygen vacancies does not have a significant effect on the thickness
of the defective domain wall, although it affects the structure. In
other words, lowering the density of oxygen vacancies we expect to
see changes in structure but no significant change in thickness.

\section{Concluding Remarks}

In this paper, we presented a semi-analytic study of the effect of
oxygen vacancies and strain on the structure of $180^{\circ}$ domain walls in
PbTiO$_3$ using a shell potential. We considered both Pb-centered
and Ti-centered domain walls with oxygen vacancies sitting on
them. We observed that Pb-centered domain walls with oxygen
vacancies are not stable (even under strain) and this is in agreement with the recent ab
initio calculations that predict Ti-centered defective domain
walls. To be able to solve for the structure of defective domain
walls semi-analytically we have to work with a high density of
oxygen vacancies on the domain wall (similar to the existing ab initio calculations). However, we believe that
density of oxygen vacancies does not have a noticeable effect on
the thickness of the defective domain wall.

We observe that oxygen vacancies change the structure of a
$180^{\circ}$ domain wall significantly. One important effect of
oxygen vacancies is that in the presence of oxygen vacancies,
displacements perpendicular to the domain wall are of the same order
of magnitude as the displacements in the tetragonal $c$-direction.
In the anharmonic lattice statics iterations we observe that
$a$-displacements have a fairly long tail, about five lattice
spacings on each side of the wall. However, the thickness of a
defective $180^{\circ}$ domain wall is about $1.5$ times that of a
perfect domain wall. This is different from the results of a recent
experimental study of $90^{\circ}$ domain walls in PbTiO$_3$ using
AFM and the observed large variations of domain wall thickness in
the internal $0.5-4.0$ nm \citep{Shilo2004}. However, our results
are in agreement with those obtained from a recent continuum model
that predicts different behaviors of $180^{\circ}$ and $90^{\circ}$
in response to point defects \citep{XiaoShenoyBhattacharya2005}. We
studied the effect of strain on both perfect and defective
$180^{\circ}$ domain walls. We observed that normal strains have a
greater effect on $a$-displacements but shear strains have a
significant effect on both $a$- and $c$-displacements. Finally, we
observe that the domain wall thickness does not change significantly
under normal or shear strains.

\section*{Acknowledgments}

We benefitted from a discussion with V. Gavini.


\end{document}